%% file: main.tex
\documentclass{article}
\usepackage{preamble}
\usepackage[a4paper, total={6.5in, 8in}]{geometry}
\usepackage{authblk}

\title{\vspace{-5em}Miss It Like Messi: \\Extracting Value from Off-Target Shots in Soccer}

\author[1]{Ethan Baron}
\author[2]{Nathan Sandholtz}
\author[3]{Devin Pleuler}
\author[1]{Timothy C. Y. Chan}

\affil[1]{University of Toronto, Toronto, Canada}
\affil[2]{Brigham Young University, Provo, Utah, United States}
\affil[3]{Maple Leaf Sports \& Entertainment, Toronto, Canada}
\date{}

\begin{document}
\maketitle

\input{abstract}

\input{introduction}

\input{data}

\input{generative_model}

\input{metric}

\input{results}

\input{conclusion}

\pagebreak

\printbibliography
\pagebreak

\appendix
\input{appendix}

\pagebreak

\end{document}

%% file: abstract.tex
\begin{abstract}
Measuring soccer shooting skill is a challenging analytics problem due to the \blue{scarcity} and highly contextual nature of scoring events. \blue{The introduction of more advanced data surrounding soccer shots has given rise to model-based metrics which better cope with these challenges. Specifically, metrics such as expected goals added, goals above expectation, and post-shot expected goals all use advanced data to offer an improvement over the classical conversion rate.}  However, all metrics developed to date assign a value of zero to off-target shots, which account for almost two-thirds of all shots, since these shots have no probability of scoring. We posit that there is non-negligible shooting skill signal contained in the trajectories of off-target shots and propose two shooting skill metrics that incorporate the signal contained in off-target shots. Specifically, we develop a player-specific generative model for shot trajectories based on a mixture of truncated bivariate Gaussian distributions. We use this generative model to compute metrics that allow us to attach non-zero value to off-target shots. We demonstrate that \blue{our proposed} metrics are more stable than current state-of-the-art metrics and have \blue{increased} predictive power.

\end{abstract}

%% file: introduction.tex
\section{Introduction}

Effectively evaluating shooting skill is a key challenge in soccer analytics. Historically, the finishing skill of players has been compared using classical statistics such as goals and shots on target, \blue{or more advanced models derived from these statistics (e.g., \cite{McHale})}. These statistics are easily obtainable but suffer from small sample sizes and a lack of comparability across players, since the probability of scoring a goal or landing a shot on-target depends heavily on the situation in which the shot was taken. Thus, a large focus in soccer analytics is the development of advanced shooting metrics that are both stable over time and comparable across players.

Much research to date has focused on the issue of comparability. For example, expected goals models address the issue of comparability by measuring a shot’s value within the context of a game situation. The \textit{pre-shot expected goals} (\xG) metric estimates the probability of a shot scoring given contextual variables at the time of the shot, such as the shot location, the proximity of other players, and the body part used (i.e., head or foot) \citep{Rathke, BrechotFlepp, Lucey2015QualityVQ, Rowlinson2020, AnzerBauer}. Subtracting a player's expected goals scored from their actual goals scored results in \textit{goals above expectation} (\GAX), which provides a measure of shooting skill that is comparable across players. Unfortunately, this metric offers limited empirical stability, as defined by \citet{MetaAnalytics}; a player's \GAX{} in one season is poorly predictive of their \GAX{} in the next season \citep{DevinEGABlog, 11tegen11}.

An improvement over \GAX{} relies on \textit{post-shot expected goals} (\PostXg), which considers the probability of a shot scoring conditional on its spatial trajectory after being struck \citep{Goodman}. \blue{Note that \PostXg{} models assign a value of zero to all off-target shots}. Subtracting a shot’s post-shot expected goals value from its pre-shot expectation results in a metric known as \textit{expected goals added} (\EGA). While the additional information utilized by \PostXg{} models has led to a more stable shooting skill metric in \EGA{} \citep{DevinEGABlog}, overall stability remains poor because the issue of limited sample size remains.

\blue{The shooting metrics currently available ignore the trajectories of off-target shots and would only be able to distinguish two different players based on their on-target shots. Since} off-target shots make up between 57\% \blue{and} 65\% of all shots \blue{in soccer} \citep{Rathke,Zhou,Mao}, \blue{these} current approaches essentially ignore most of the available data. In a sport that already suffers from limited sample sizes, uncovering a signal from off-target shots has the potential to significantly improve shooting metrics.

In basketball, \citet{RBFG} showed that using the trajectories of all shots, not just on-target shots, leads to an improved estimate of player field goal percentage. In soccer, \citet{DevinFreeKickBlog} shows that including the trajectory information of all of a player's free kicks, including off-target shots, yields more stable estimates of shooter skill. We hypothesize that there is a non-negligible shooting skill signal contained in the trajectories of off-target shots. Intuitively, a player with many shots that narrowly miss an upper corner of the goal is likely a better shooter than one who consistently misses wildly. \blue{Therefore, identifying players who have many near misses as opposed to wild misses can help determine shooting skill.}

In this paper, we develop a hierarchical generative model for the trajectories of soccer shots. We use this generative model to propose \blue{two} novel shooting skill metrics that extract signal from the trajectories of all shots, including off-target shots. We show that our proposed metrics are more stable over time, helping to address the issue of small sample sizes. We also verify that our metrics carry predictive value of future player performance and therefore contain a valuable signal about player shooting skill.

The paper is structured as follows. In Section \ref{Data}, we describe our data. In Section \ref{GenModel}, we develop our generative model for shots. In Section \ref{PropMetrics}, we explain how we use this generative model to develop our novel shooting skill metrics. In Section \ref{Results}, we validate our proposed metrics and demonstrate their improved stability over existing metrics. We conclude in Section \ref{Conclusion}.

%% file: data.tex
\section{Data}
\label{Data}

We use data on \blue{77,315} shots from StatsBomb Services Ltd, obtained via an academic partnership with Toronto FC. The data format is event data from six international leagues, comprising 15 total seasons. The available leagues range in ranking from 7 to 31, \blue{and therefore represent a comparable level of competition, as none of these are elite leagues but all of them are home to high-level professional players. As a concrete example, the mean pre-shot expected goals value for each league and season has a standard deviation of about 0.0015}. See Table \ref{LeaguesTable} for a summary.

\begin{table}[ht!]
\centering
\begin{tabular}{|l|l|c|l|c|}
\hline
League                                      & Countries                                   & Ranking              & Season  & \# Shots \\ \hline
\multirow{3}{*}{Eredivisie}                 & \multirow{3}{*}{Netherlands}              & \multirow{3}{*}{7}  & 2018-19 & \blue{5586}     \\ \cline{4-5} 
                                            &                                           &                     & 2019-20 & \blue{4102}     \\ \cline{4-5} 
                                            &                                           &                     & 2020-21 & \blue{5060}     \\ \hline
\multirow{3}{*}{Major League Soccer (MLS)}  & \multirow{3}{*}{United States, Canada} & \multirow{3}{*}{14} & 2018    & \blue{7056}     \\ \cline{4-5} 
                                            &                                           &                     & 2019    & \blue{7435}     \\ \cline{4-5} 
                                            &                                           &                     & 2020    & \blue{5297}     \\ \hline
Argentine Primera División                  & Argentina                                 & 18                  & 2019-20 & \blue{4425}     \\ \hline
\multirow{3}{*}{2. Bundesliga}              & \multirow{3}{*}{Germany}                  & \multirow{3}{*}{20} & 2018-19 & \blue{5123}     \\ \cline{4-5} 
                                            &                                           &                     & 2019-20 & \blue{5213}     \\ \cline{4-5} 
                                            &                                           &                     & 2020-21 & \blue{4934}     \\ \hline
\multirow{3}{*}{Ligue 2}                    & \multirow{3}{*}{France}                   & \multirow{3}{*}{26} & 2018-19 & \blue{5534}     \\ \cline{4-5} 
                                            &                                           &                     & 2019-20 & \blue{4190}     \\ \cline{4-5} 
                                            &                                           &                     & 2020-21 & \blue{5457}     \\ \hline
\multirow{2}{*}{United Soccer League (USL)} & \multirow{2}{*}{United States, Canada} & \multirow{2}{*}{31} & 2019    & \blue{3273}     \\ \cline{4-5} 
                                            &                                           &                     & 2020    & \blue{4630}     \\ \hline
\end{tabular}
\caption[Available Leagues]{Leagues and seasons included in our dataset. Rankings are per \citet{LeagueRankings} on January 23, 2022.}
\label{LeaguesTable}
\end{table}

We use the following features of each shot in our analysis:
\begin{itemize}
    \item player who attempted the shot
    \item shot outcome (Saved, Goal, Off Target, Blocked, or Post)
    \item coordinates of the shot start $(x, y)$ and end $(x, y, z)$ locations
    \item body part used for the shot (Right Foot, Left Foot, Header, or Other)
    \item estimated \xG{} and \PSXG{} for the shot, based on StatsBomb-developed models
\end{itemize}

\begin{figure}[ht!]
\centering
\includegraphics[width=\textwidth]{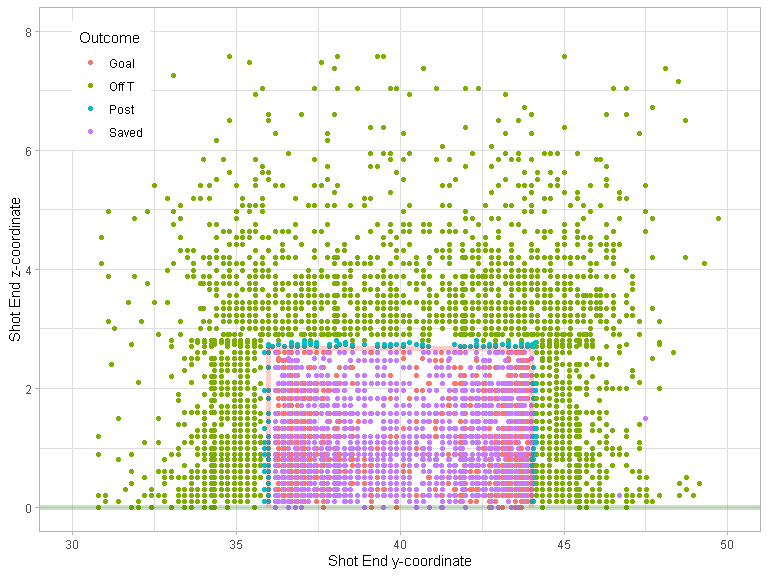}
\caption[Processed 2018 MLS Data]{Adjusted shot end coordinates from 2018 MLS data, \blue{coloured by outcome}. Note that each point may represent multiple shots, as the data are collected on a discrete grid.}
\label{PreprocessedMLS2018}
\end{figure}

Figure \ref{PreprocessedMLS2018} shows the shot end coordinates for 2018 MLS data. We perform several preprocessing steps with the data. First, we made a correction to fix a bias in the data collection process affecting shot end coordinates near the goal frame (see Appendix \ref{CorrectingCoordinates} for details). Second, for saved and blocked shots, we project forward the shot trajectories to estimate where they would have crossed the goal line had they not been obstructed. We describe this imputation process below.
\blue{Third, we exclude shots taken from closer than 6 yards to the goal, since such shots often demand force and luck rather than accuracy. Finally, to exploit symmetry in the data, we reflect the $y$ coordinates of left-footed shots about the middle of the goal.}

\subsection{Imputing Shot Trajectories}
For most our data, the reported shot end coordinates for saved \blue{or blocked} shots are the coordinates where the shot was \blue{obstructed}, rather than the coordinates where the shot would have crossed the goal line had it not been \blue{obstructed}.  \blue{We project the trajectories of these shots forward to determine the $(y, z)$ coordinates where they} would have crossed the goal line had they continued in their \blue{paths}.

\blue{For the end $y$ coordinate, we project the horizontal path of the ball linearly and compute the $y$ coordinate where the trajectory intersects with the goal line. For the $z$ coordinate, a physics-based model can be used to project the flight path and determine the height at which the shot would cross the goal line. However, such a model would rely on reliable shot timestamps for the shot start and end and a model for the shot bouncing. We found that the shot durations in the data had higher-than-expected variance, leading to noisy estimates of the shot end $z$ coordinates. We therefore chose to simply use the provided $z$ coordinate of the save as a proxy for the projected end $z$ coordinate.}

\begin{figure}[!ht]
    \centering
    \subfloat[\centering Original]{\includegraphics[width=0.47 \textwidth]{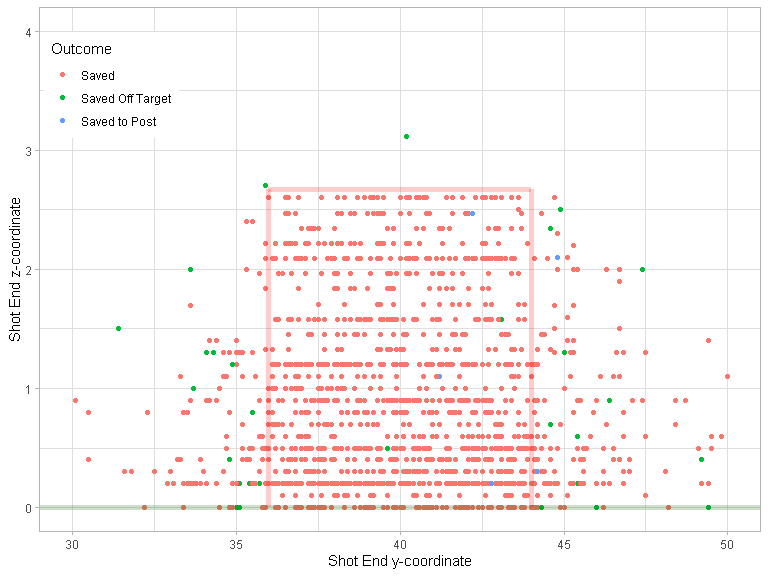}}
    \qquad
    \subfloat[\centering Projected]{\includegraphics[width=0.47 \textwidth]{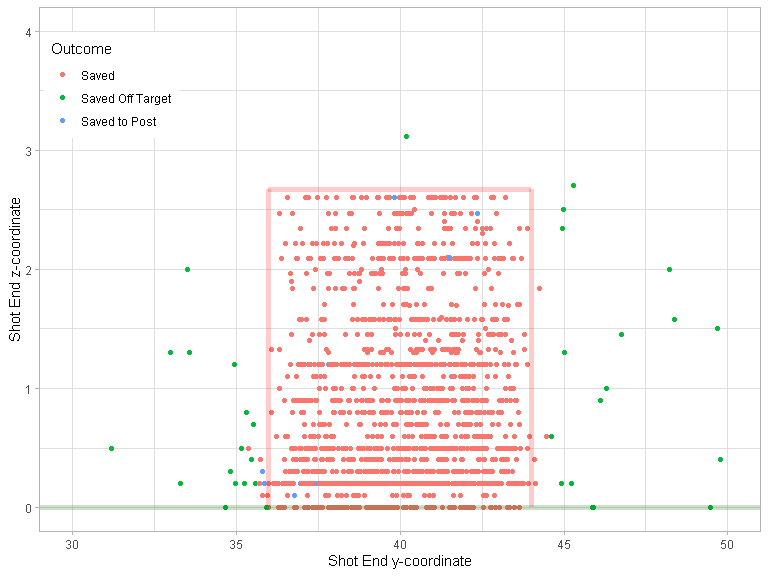}}
    \caption[Projected Shot End Coordinates for 2020 USL Data]{Comparison of original and projected shot end coordinates for saved shots from 2020 USL data.}
    \label{ProjectedCoordinatesComparison}
\end{figure}

Figure \ref{ProjectedCoordinatesComparison} shows a comparison of the original and projected shot end coordinates for saved shots in the 2020 USL data. Note that the projected coordinates effectively distinguish between on-target and off-target saved shots \blue{as labelled by StatsBomb}, with only a few exceptions.

\subsection{Experimental Execution Error}

\citet{Hunter} performed an experiment to record the accuracy of penalty shots taken by semi-professional soccer players aimed at either a high (1.75 yards above the ground) or a low (0.14 yards above the ground) target location. They collected data on the two-dimensional accuracy of 8466 shots from 21 players. Subjects were asked to attempt shots with both left and right feet, using different shot techniques, and with slower or faster intended speeds. We utilize this data to inform the covariance specification of \blue{the} generative shot model in the next section.  While the execution error patterns observed in \blue{\citet{Hunter} may be different than those of professional} players in real match situations, using this experimental data allows us to make informed assumptions about the shape of the execution errors of soccer shots from real matches.

%% file: generative_model.tex
\section{Generative Model for Shots}
\label{GenModel}

The metrics we propose rely on a generative model for soccer shots. Specifically, we model the shot end coordinates, denoted $(y, z)$, corresponding to the location where the ball crosses the plane of the goal line. We model $(y,z)$ using a mixture of bivariate Gaussian distributions, truncated so that the $z$-coordinates are non-negative. The choice of bivariate Gaussians is informed by empirical observations of variability in soccer shots \citep{Hunter}. Bivariate Gaussian distributions have also been used in similar models in other sports, including darts \citep{Darts, HaughDarts} and tennis \citep{CraigTennis}. \blue{To share information across players, we include a hierarchical component whereby each player's mixture model parameters are shrunk towards a global prior.}

For shot $i$ from player $p$, we model the probability density function for the shot's end $(y,z)$ coordinates with the following hierarchical mixture model:
\begin{align}
    f(y_{i}^{(p)}, z_{i}^{(p)}) &= \sum_{k = 1}^K \theta^{(p)}_{k} \cdot \textrm{TruncNorm}(y_{i}^{(p)}, z_{i}^{(p)} | \boldsymbol{\mu}_k, \boldsymbol{\Sigma}_k) \label{eq:initial_likelihood} \\
    \boldsymbol\theta^{(p)} &\sim \text {Dirichlet}_{K}(\alpha \cdot \boldsymbol\beta) \label{eq:hierarchy}
\end{align}
Here, $\boldsymbol{\mu}_k$ and $\boldsymbol{\Sigma}_k$ are the mean vector and covariance matrix respectively of the $k${th} mixture component, $K$ is the number of mixture components, $\theta^{(p)}_{k}$ is the $k$th mixture weight for player $p$, and $\boldsymbol\beta = \{\boldsymbol\beta_{1}, \ldots, \boldsymbol\beta_{K}$\} is the vector of global component weights (i.e., average across all players). The parameter $\alpha$ is a hyperprior that governs the amount of player shrinkage toward the global component weights.  Note that $\sum_{k = 1}^K \boldsymbol\beta_{k} = 1$ and $\sum_{k = 1}^K \theta^{(p)}_{k} = 1 \text{ for all } p$. 

Ideally, we would specify priors for the component parameters \blue{$\boldsymbol{\mu}_k$ and $\boldsymbol{\Sigma}_k$} and fit the mixture via a Bayesian inference framework. For example, we might use a uniform distribution over the goal mouth as a prior for the component means. Unfortunately, directly estimating a mixture of truncated bivariate Gaussians is computationally challenging, as acceptance rates for the components must be repeatedly recomputed using sampling for each new mean and covariance \citep{Wilhelm_Manjunath_2010}. \citet{TruncatedGaussianMixture} provide an expectation-maximization (EM) algorithm to estimate mixtures of truncated multivariate Gaussians in MATLAB, but such estimation within a Bayesian hierarchical framework has not been implemented. In our case, the number of components is also unknown \emph{a priori}, introducing additional complexity.

To cope with these computational challenges, we fit \blue{the generative} model in two stages. First, we create a dense set of fixed mixture components and focus our inference on the \blue{global} component weights. This allows us to then use a pruning heuristic over the components \blue{by removing low-weight components, yielding} a more parsimonious representation. In the second stage, we estimate player-specific weights over the resulting components via a Bayesian hierarchical model. We describe these two stages \blue{in more detail} below.

\subsection{Global Mixture Model Components}

Following \citet{JeffreysPriors}, we first saturate the mixture model with a large number of fixed components and then prune components with low weights. Intuitively, this large set of fixed components represents a coarse parameter space over which we will make inference on the components.  

To do this, we first create a dense set $\mathcal{M} = \{\mathbf{m}_1, \ldots, \mathbf{m}_J\}$ of evenly spaced points within the goal frame representing the set of component means. Intuitively, these points correspond to possible intended shot locations. Further, we want the saturated set of components to span a range of execution error scales (i.e., covariance matrices) around the specified locations.  We therefore enumerate multiple Gaussians with different covariance matrices at each location $\mathbf{m}_j \in \mathcal{M}$.  We associate each location with a nominal covariance matrix $\mathbf{S}_j$, but we multiply $\mathbf{S}_j$ by a varying scaling factor $\lambda_\ell$, $\ell = 1,\ldots,L$, yielding $L$ unique covariance matrices for each component mean.  In essence, this yields a set of $J \times L$ components that define the inference space on the mixture components $\boldsymbol\mu_k$ and $\boldsymbol{\Sigma}_k$.

\blue{Our model calls for the specification of the hyperparameters $M$, $\mathbf{S}_j$, and the $\lambda_\ell$.} To construct $\mathcal{M}$, we experimented with two parameters, $n_y$ and $n_z$, denoting the number of components in the $y$ and $z$ direction, respectively. Given $n_y$ and $n_z$, the component locations were placed equidistantly between the edges of the goal frame. We tested values for $n_y \in [4, 11]$ and $n_z \in [3, 8]$. We use empirical observations of execution error in soccer from \citet{Hunter} to inform the shape of our covariance matrices $\mathbf{S}_j$ for each location $\mathbf{m}_j$. See Appendix \ref{FittingCovariances} for a description of how the $\mathbf{S}_j$ were obtained. Finally, for the $\lambda_\ell$, we performed a random search, using weakly informative normal distributions to generate candidate $\lambda_\ell$ for each $L \in \{2, 3\}$.

\blue{After comparing the performance from 30 combinations of hyperparameter settings, we observed that the specific choice of hyperparameters had a relatively small impact on our results. The most notable exception is that the performance seemed to improve with larger $n_y$ and larger $n_z$ up until $n_z = 6$. Our final selected hyperparameters were $n_y = 11$, $n_z = 6$, $L = 2$, $\lambda_1 = 1.0$, and $\lambda_2 = 3.8$. Therefore, our saturated mixture model includes $11 \times 6 \times 2 = 132$ components layered onto $11 \times 6 = 66$ distinct mean locations $\mathbf{m_j}$.}

Having created this dense set of components, we fit the following saturated model to all shot end coordinates via Bayesian inference.
\begin{align}
f(y_{i}^{(p)}, z_{i}^{(p)}) &= \sum_{j \in J} \sum_{\ell \in L} \boldsymbol\beta_{j\ell} \cdot \textrm{TruncNorm}(\mathbf{m}_j, \lambda_\ell \mathbf{S}_j). \\
\boldsymbol\beta &\sim \text {Symmetric-Dirichlet}_{(J\times L)}(\alpha = 1/2)    
\end{align}
Following \cite{Rousseau}, we place a symmetric Dirichlet prior on the global mixture model component weights $\boldsymbol\beta_{ij}$. This choice of prior is neutral in that no component is favored over the others. Additionally, setting $\alpha = 1/2$ for the symmetric Dirichlet parameter corresponds to the Jeffreys prior associated with this problem \citep{JeffreysPriors}, and is therefore uninformative. This choice of $\alpha = 1/2$ encourages the posterior to be conservative in the number of components, driving the weights of superfluous components towards zero \citep{Rousseau}, and allowing us to identify a subset of the components that drive the model.

\begin{figure}[!bht]
\centering
\includegraphics[width= 0.9 \textwidth]{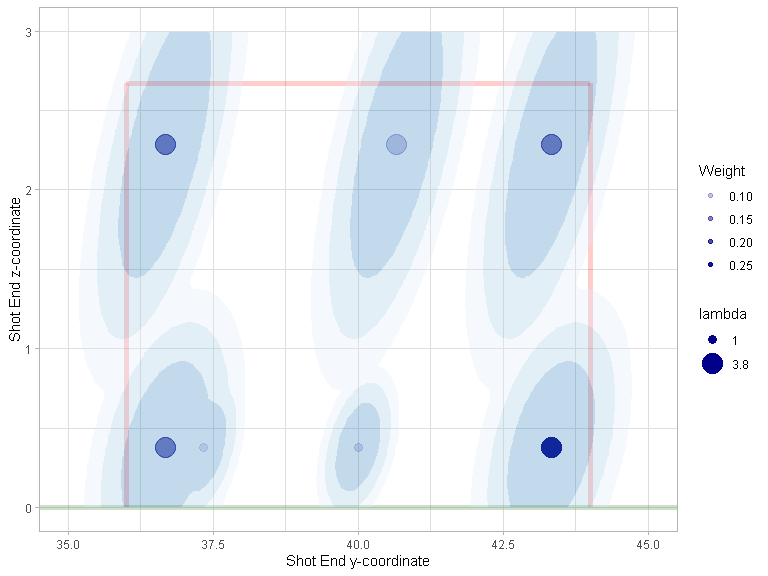}
\caption[Shot Generative Model]{Means and covariances of components used in trimmed shot generative model. The size of the circles \blue{corresponds to the value of $\lambda$ associated with the component, while the opacity reflects the weight associated with the component}.}
\label{TrimmedModel}
\end{figure}

Due to the computational burden of fitting the full hierarchical model on the saturated set of components, the only parameters we estimate at this stage are the global component weights, $\boldsymbol\beta$. Upon estimating the posteriors for the global component weights using this symmetric Dirichlet prior, we then prune components that have estimated weights below a small threshold, as suggested by \cite{Rousseau}. Specifically, we eliminate from the model all components with $\hat{\boldsymbol\beta}_{j\ell} < 0.01$, where $\hat{\boldsymbol\beta}_{j\ell}$ is the mean of the $(j,\ell)$ mixture weight posterior. This yields a small subset of the components in the saturated regime that drive the model, which we denote as $C$. We then re-estimate $\boldsymbol\beta$ after selecting the trimmed set of mixture components. The \blue{selected subset of} components, along with their relative weights, are illustrated in Figure \ref{TrimmedModel}.

\subsection{Hierarchical Mixture Model} \label{sec:HMM}

Having estimated the mixture components, we can re-express the likelihood from \eqref{eq:initial_likelihood} as:
\begin{align}
    f(y_{i}^{(p)}, z_{i}^{(p)}) &= \sum_{k = 1}^K \theta^{(p)}_{k} \cdot \textrm{TruncNorm}(y_{i}^{(p)}, z_{i}^{(p)} | \mathbf{m}_k, \mathbf{S}_k) \label{likelihood}
\end{align}
where $(\mathbf{m}_k, \mathbf{S}_k)$ is the $k$th element of $C$ and $K = |C|$.

We now seek to estimate player-specific weights over these components via the hierarchical framework in \eqref{eq:hierarchy}. \blue{We use $\alpha=30$ after comparing the mean log-likelihood per shot for several values of $\alpha$}. We fit the models using variational inference in RStan \citep{RSTan}, as \blue{MCMC} sampling resulted in prohibitively long computation times.

%% file: metric.tex
\section{Proposed Shooting Metrics}
\label{PropMetrics}

\blue{In this section, we present our two proposed shooting metrics. The first, which we call RBPostXg, reflects a player's theoretical probability of scoring based on the generative model described above. The second, which we call GenPostXg, is a more flexible shot-specific alternative to RBPostXg.}

\subsection{\blue{RBPostXg: Rao-Blackwellized PostXg}}

\blue{The naming of this metric follows the precedent set in basketball by \citet{RBFG}. They propose RB-FG\%, an estimate of a player's field goal percentage based on shot trajectory information beyond whether the shot was made or missed. Using regression, they estimate each shot's make-probability given this additional information.  They then condition on these estimated probabilities in their proposed metric.  We apply a similar process in our work by conditioning our metric on player-specific parameter estimates underlying the generative model for their shot trajectories.}  

\blue{Our first proposed shooting metric is simply the probability of a given player's shot resulting in a goal, conditioned on the parameter estimates underlying that player’s generative model.}
Let $\{x^{(p)}_1, \ldots, x^{(p)}_{N_p}\}$ be the observed shots from player $p$, with $x^{(p)}_i$ equaling 1 if shot $i$ resulted in a goal and 0 otherwise. Let $\{(y^{(p)}_1, z^{(p)}_1), \ldots, (y^{(p)}_{N_p}, z^{(p)}_{N_p})\}$ represent the end coordinates of these shots.
We assume that the trajectories of a player's shots are i.i.d. draws from the shot-generating process described above. For player $p$, the probability of a shot being sampled from component $(\mathbf{m}_k, \mathbf{S}_k) \in C$ is given by $\theta^{(p)}_k$. We also assume that the probability of a shot resulting in a goal given its end coordinates is $\textrm{\PostXg{}}(y,z)$, essentially a simple post-shot expected goals model which we develop and describe below. Therefore, given $(y^{(p)}_i, z^{(p)}_i)$, we can represent the unobserved shot outcome as a Bernoulli random variable:
\begin{align}
    \mathbf{X}^{(p)} \, | \, (y^{(p)}_i, z^{(p)}_i) &\sim \textrm{Bernoulli}(\textrm{\PSXG{}}(y^{(p)}_i, z^{(p)}_i)). 
\end{align}
We wish to estimate $\mathbb{E}[\mathbf{X}^{(p)} | \mathbf{\boldsymbol\theta^{(p)}}]$, the probability that an arbitrary shot from player $p$ results in a goal.  We have:
\begin{align}
    \mathbb{E}[\mathbf{X}^{(p)} | \mathbf{\boldsymbol\theta^{(p)}}] &= \int_y \int_z \textrm{\PSXG{}}(y, z) \cdot \sum_{k = 1}^K \theta^{(p)}_{k} \cdot \textrm{TruncNorm}(\mathbf{m}_k, \mathbf{S}_k) \, dy \, dz \\
    &= \sum_{k = 1}^K \theta^{(p)}_{k} \int_y \int_z \textrm{\PSXG{}}(y, z) \cdot \textrm{TruncNorm}(\mathbf{m}_k, \mathbf{S}_k) \, dy \, dz \\
    &= \sum_{k = 1}^K \theta^{(p)}_{k} \cdot v_{k}
\end{align}
where 
\begin{equation}
    v_{k} := \int_y \int_z \textrm{\PSXG{}}(y, z) \cdot \textrm{TruncNorm}(\mathbf{m}_k, \mathbf{S}_k) \, dy \, dz
\end{equation}
and all other terms are as defined in Section \ref{sec:HMM}.  
These $v_{k}$ represent the mean expected goals value for a shot sampled from the $k$th component of the mixture model. We \blue{estimate} these numerically using Monte Carlo integration and rejection sampling from untruncated bivariate Gaussian distributions. 

\blue{We define \textit{\RB{}ized \PSXG{}} as}
\begin{equation} \textrm{\RBPostXg}(p) := \sum_{k = 1}^K \hat{\theta}^{(p)}_{k} \cdot \hat{v}_{k} \end{equation}
\blue{which is the predicted probability that an arbitrary shot from player $p$ results in a goal.}

\subsection{\blue{GenPostXg: Shot-Specific Generalization of RBPostXg}}
\blue{In the previous section, our metric leverages off-target shot information indirectly by its dependence on $\boldsymbol{\theta}^{(p)}$, but it does not explicitly attach value to individual shots.  In this section, we create a metric that actually assigns non-zero value to every shot in the dataset, including off-target shots.  In the context of our framework, since we treat each shot trajectory as a random draw from a generative model, if the shot were taken again in the exact same situation, the outcome could be different.  In our second metric, we attach value to shots by marginalizing over all the potential shots that could have been observed, which will invariably include some on-target shots.}

Specifically, for a  shot with end coordinates $(y,z)$, we compute the probability $p_{k} (y,z)$ that the shot was generated by the $k$th mixture component as:
\begin{equation} \hat{p}_{k} (y,z) = \frac{\hat{\beta}_{k} \cdot \textrm{TruncNorm}(y, z | \mathbf{m}_k, \mathbf{S}_k)}{\sum_{k = 1}^K \hat{\boldsymbol\beta}_{k} \cdot  \textrm{TruncNorm}(y, z | \mathbf{m}_k, \mathbf{S}_k)} \end{equation}
Note that here we use the \blue{estimated} global component weights $\hat{\boldsymbol \beta}$ for all shots.  We then define \emph{Generalized \PSXG{}} as:
\begin{equation}\textrm{GenPostXg}(y,z) = \sum_{k = 1}^K  \hat{v}_{k} \cdot \hat{p}_{k}(y, z).\end{equation}
\blue{A player's GenPostXg for a given time period is defined as the mean GenPostXg value of their individual shots.}

\subsection{Coordinates-based \PSXG{} Model}

\blue{Both our proposed metrics} rely on the function $\textrm{\PSXG{}}(y, z)$, a \PSXG{} contour over the frame of the goal mouth. Although StatsBomb provides estimates for the \PSXG{} value of individual shots, these estimates are based on a proprietary model that includes many attributes beyond the shot end coordinates. Therefore, we develop a \PSXG{} model relying only on shot end $(y,z)$ coordinates. Specifically, we model the log-odds of scoring as a polynomial function of the shot end $(y,z)$ coordinates using a logistic regression formulated as follows:\footnote{The $\beta$s in this equation represent different parameters than those defined earlier.  We felt this abuse of notation was warranted as regression models usually use this notation for the coefficients.}

\begin{equation}\log{\frac{\textrm{\PSXG{}}(y, z)}{1-\textrm{\PSXG{}}(y, z)}} = \beta_0 + \beta_1 y + \beta_2 y^2 + \beta_3 y^3 + \beta_4 z + \beta_5 z^2 + \beta_6 z^3. \end{equation}

\begin{figure}[!ht]
\centering
\includegraphics[width=0.9\textwidth]{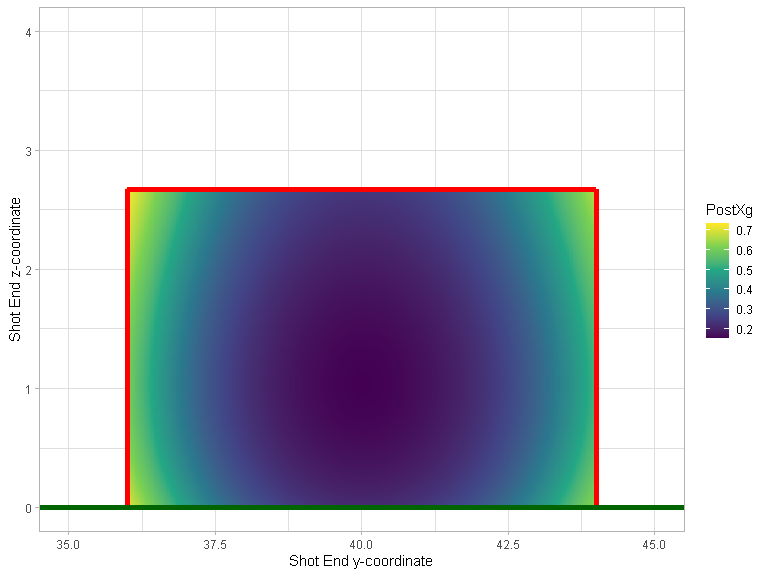}
\caption[\PSXG{} Contour]{Contour of \PSXG{}(y, z) model. \blue{Regions where a shot is more likely to score are shown in brighter colours, whereas regions where a shot is less likely to score are shown in darker colours. Note that all shot end coordinates outside the goal frame have a PostXg value of zero.}}
\label{PostXgContour}
\end{figure}

Our model results in the intuitive \PostXg{} contour for on-target shots shown in Figure \ref{PostXgContour}. The area under the receiver-operator curve for our model is \blue{0.70, compared to 0.86 for the StatsBomb-provided PostXg model}, which considers additional attributes such as the shot speed and goalkeeper location.

%% file: results.tex
\section{Results}
\label{Results}

\blue{In this section, we demonstrate some key results that emerge from our proposed metrics and demonstrate how our approach can be used in practice to analyze player shooting patterns. 
Although our data cannot be shared publicly, the code used to implement our methods and generate these results is made open-source at} \url{https://github.com/baronet2/shotmissr}.

\subsection{\blue{Stability and Predictive Value}}

\blue{In order for our proposed metrics to offer value to teams, they should be predictive of future player value. Due to the limited number of consecutive seasons available to us in the data, we measure this by considering the correlation between a player's performance on our metrics in the first and second halves of a given season.}

\blue{Table  \ref{InterseasonCorrelationsTable} reports the correlation in the values of metrics from the first half-seasons to the second half-seasons, comparing our proposed metrics to existing metrics. The correlation between player performances using the same metric in both half seasons provides a measure of the stability of that metric. For the benchmark metrics, the stability is very low, reaching at most 0.056. Both of our proposed metrics offer an increase in stability, up to 0.136 for RBPostXg and 0.162 for GenPostXg. Furthermore, our proposed metrics, and especially GenPostXg, are better predictors of future performance on the benchmark metrics than these benchmark metrics themselves, confirming that our proposed metrics are indeed capturing a signal in player skill.}

\begin{table}[ht!]
\centering
\begin{tabular}{l|rrrrr}
\hline
& \multicolumn{4}{c}{Second half-season} \\
First half-season &  GAX &    EGA &  \textit{RBPostXg} &  \textit{GenPostXg} \\
\hline
GAX       &  0.035 &  0.026 &     0.006 &      0.024 \\
EGA       &  0.025 &  0.056 &     0.003 &      0.010 \\
\textit{RBPostXg}  &  0.032 &  0.042 &     \textbf{0.136} &      0.141 \\
\textit{GenPostXg} & \textbf{0.074} &  \textbf{0.072} &     0.132 &      \textbf{0.162} \\
\hline
\end{tabular}
\caption[Inter-season Correlations]{\blue{Inter-season correlation of metrics from first to second half-season. For example, the correlation between players' GAX in the first half of a season to their EGA in the second half of a season is 0.026. The italicized metric names are our proposed new metrics. The best predictor of each metric in the second half of the season is bolded.}}
\label{InterseasonCorrelationsTable}
\end{table}

Table \ref{InterseasonCorrelationsTable} includes the metrics for all players, including those with only a few shots in the database, which make up a large component of the data. To account for the potential \blue{bias} introduced by including low sample-size players, Table \ref{CorrelationsTable40Plus} reports similar values but including only the player-seasons with at least 40 shots.
Here, the stability of the \blue{benchmark metrics remains low, even dipping below zero. On the other hand}, our proposed metrics all offer significantly higher stability, up to \blue{0.232} in the case of GenPostXg. This is an indication that our metrics are capturing a consistent signal from players' shot end coordinates. \blue{Again, our proposed metrics offer the highest predictive power for future performance on the benchmark metrics, demonstrating that our metrics contain a true predictive signal for player shooting skill.}

\begin{table}[ht!]
\centering
\begin{tabular}{l|rrrrr}
\hline
& \multicolumn{4}{c}{Second half-season} \\
First half-season &  GAX &    EGA &  \textit{RBPostXg} &  \textit{GenPostXg} \\
\hline
GAX       &  -0.025 &  -0.065 &     -0.040 &      -0.023 \\
EGA       &  -0.025 &  -0.033 &     -0.053 &      -0.076 \\
\textit{RBPostXg}  &  0.028 &  \textbf{0.035} &   \textbf{0.219} &      0.226 \\
\textit{GenPostXg} & \textbf{0.037} &  0.020 &     \textbf{0.219} &      \textbf{0.232} \\
\hline
\end{tabular}
\caption[Inter-season Correlations]{\blue{Inter-season correlation of metrics from first to second half-season for player-seasons with 40 or more shots. This table presents the same information as Table \ref{InterseasonCorrelationsTable} after filtering to player-seasons with at least 40 shots. The italicized metric names are our proposed new metrics. The best predictor of each metric in the second half of the season is bolded.}}
\label{CorrelationsTable40Plus}
\end{table}

\begin{figure}[ht!]
\centering
\includegraphics[width=\textwidth]{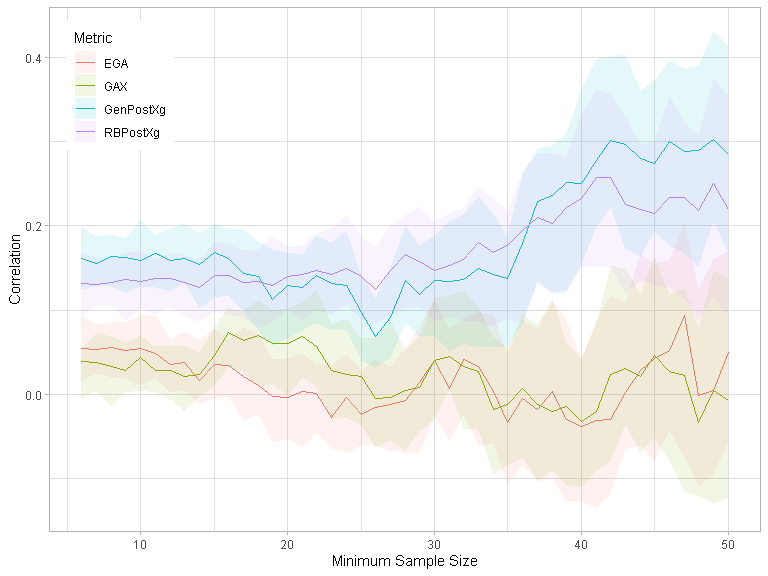}
\caption[Inter-season correlation of metrics as function of player sample size threshold]{\blue{Inter-season correlation of metrics as function of player sample size threshold. The sleeves show 90\% confidence intervals estimated via bootstrapping.}}
\label{InterseasonCorrelationsPlot}
\end{figure}

Figure \ref{InterseasonCorrelationsPlot} plots these inter-season correlations \blue{of each} metric for the first and second half-seasons as a function of the sample size threshold for shots in a player-season (e.g., \blue{40} for Table \ref{CorrelationsTable40Plus}). We see that \blue{our proposed metrics outperform} \GAX{} and \EGA{} for all sample size thresholds. \blue{Our metrics reach} an inter-season correlation of almost \blue{0.3 for a sample size threshold of 40 shots or more, whereas \GAX{} and \EGA{} continue to have almost no inter-season stability even for higher sample size thresholds}. This figure demonstrates that \blue{our proposed metrics identify} reliable estimates of player shooting skill faster than previous metrics. In other words, by including information extracted from the trajectories of off-target shots, \blue{RBPostXg and GenPostXg are} able to identify signal of shooting skill even \blue{with small} sample sizes.

The result shown in Figure \ref{InterseasonCorrelationsPlot} is particularly important if one considers how soccer analysts would use a metric like GenPostXg to evaluate talent. If \blue{a team can reliably identify skilled shooters earlier than their opposition, that team would be able to recruit higher-quality players more cheaply, leading to greater success on the pitch and higher profitability in the transfer market.}

\blue{Recall that in our data processing, we filtered the shots in our dataset according to the shot distance, and reflected the coordinates of left-footed shots. A sensitivity analysis with respect to these choices is presented in Appendix \ref{SensitivityAnalysis}. The sensitivity analysis confirms that our results are robust to the exact data preprocessing steps.}

\subsection{\blue{Player Analysis}}

\blue{By examining the mixture model component weights of individual players, analysts can exploit information about the shooting habits of certain players.}

\begin{figure}[!ht]
\centering
\includegraphics[width=\textwidth]{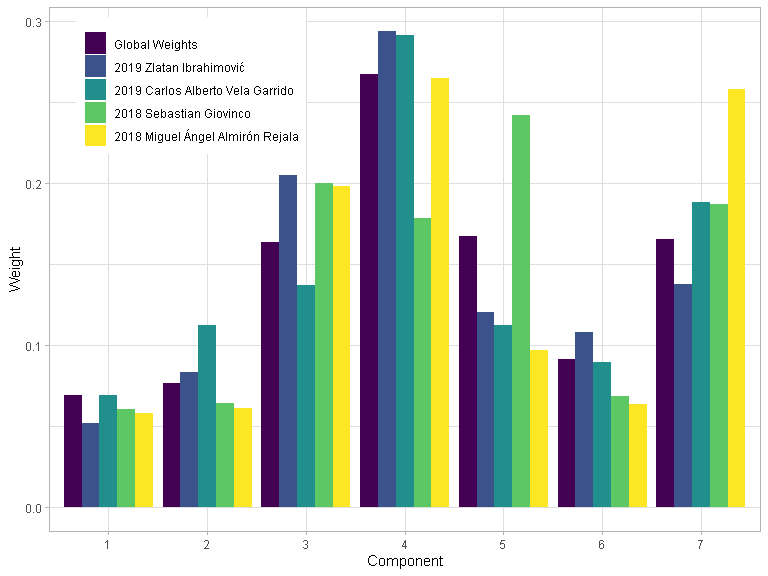}
\caption{\blue{Component weights for selected player-seasons. The x-axis indexes the components of the mixture model, and the order is arbitrary. The global mixture model weights are shown in the leftmost bar of each group, in purple.}}
\label{ComponentWeights}
\end{figure}

\blue{Figure \ref{ComponentWeights} shows the component weights fit to the first half-season of data for selected player-seasons alongside the global weights estimated on the entire data. Note that although the player component weights generally follow the patterns in the global weights, there are significant fluctuations between the different players.}

\blue{Examining the component weights for a player of interest can reveal useful information about their shooting habits. For example, in 2018, Sebastian Giovinco had an unusually high weight of around 0.24 for the fifth component. This component corresponds to the component nearest to the top-left corner of the goal. An opposing team's coach could inform their team of this information, enabling the defenders and goalkeeper to better anticipate the direction of Giovinco's shots.}

\blue{Additionally, the first and third components correspond to the two components located near the bottom-left corner of the goal with the smaller and larger values of $\lambda$ respectively. Due to the smaller variance, the first component offers a 28\% higher value compared to the third component. When considering his 2019 performance, Zlatan Ibrahimovic had a higher-than-average weight for the third component and lower-than average weight for the first component. This implies that he was less accurate than the average player when aiming at the bottom-left corner of the goal. A coach could use this information to prescribe training for Ibrahimovic to improve his accuracy when aiming for the bottom-left corner.}

%% file: conclusion.tex
\section{Conclusion}
\label{Conclusion}

In this paper, we develop a \blue{hierarchical generative model} for soccer \blue{shot coordinates} based on a mixture of truncated bivariate Gaussian distributions. \blue{The player parameters from our hierarchical generative model allow teams to analyze specific players' shooting habits and diagnose potential areas for improvement or prepare for an upcoming opponent. Furthermore, we} use this shot-generating process to develop \blue{two} novel metrics that extract value from off-target shot trajectories. We \blue{show} that our proposed metrics are significantly more stable than previous state-of-the-art metrics, and exhibit improved predictive value of a player’s performance in the second half of a season given their shots in the first half of the season. Our \blue{proposed metrics are valuable in recruitment and player valuation settings as they allow} analysts to obtain more reliable estimates of shooting skill in players, even given limited sample sizes.

%% file: appendix.tex
\section{Appendix}

\subsection{Adjusting Shot End Coordinates}
\label{CorrectingCoordinates}

\begin{figure}[h!]
\centering
\includegraphics[width=\textwidth]{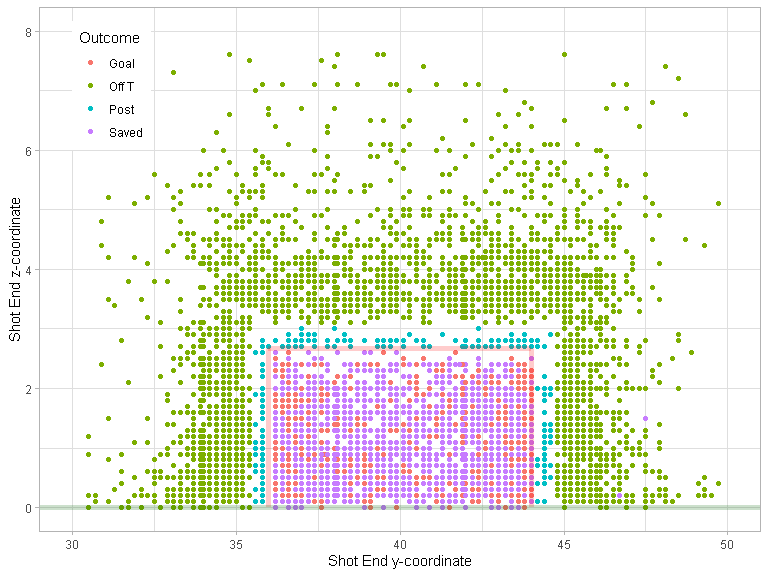}
\caption[Original 2018 MLS Data]{Original shot end coordinates for 2018 MLS data. \blue{Shots are coloured by their outcome. Note the significantly lower density of shots in the regions just outside the goal frame shown in red.}}
\label{OriginalMLS2018}
\end{figure}

In our data pre-processing, we first applied a small correction to fix an apparent bias in the data collection process affecting shot end coordinates near the goal frame. For a few seasons of data (MLS 2018, Ligue 2 2018-19, 2. Bundesliga 2018-19, Eredivisie 2018-19), the density of shots with end coordinates just outside the posts was low relative to the surrounding regions (see Figure \ref{OriginalMLS2018}). We hypothesize that this effect was caused by StatsBomb's data collection system, which amplifies the size of the posts so that data collectors are can click on them more easily. To correct for this effect, shot end coordinates were shifted to preserve the actual width of the posts, thereby resulting in a more continuous density of shot end locations. Figure \ref{PreprocessedMLS2018} shows the distribution of shot end coordinates for the 2018 MLS season after this correction.

\subsection{Estimating Execution Error}
\label{FittingCovariances}

\begin{figure}[!htbp]
\centering
\subfloat{\includegraphics[width=0.7\textwidth]{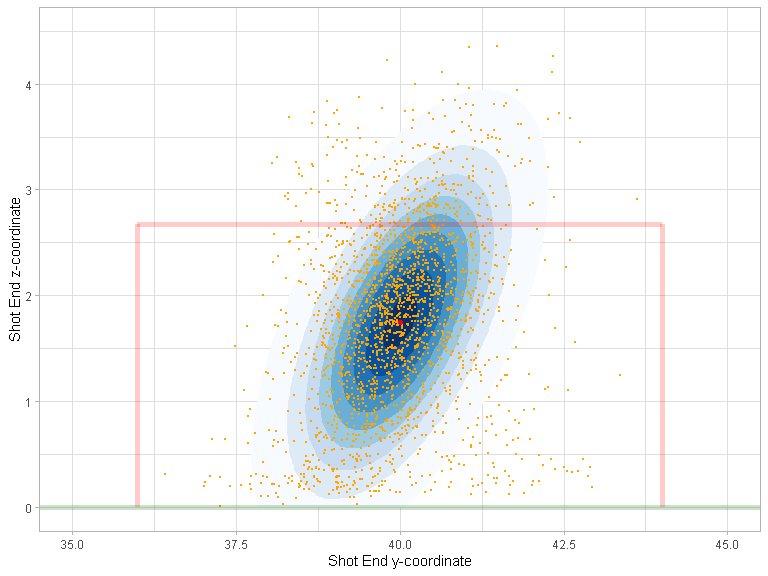}}
\qquad
\subfloat{\includegraphics[width=0.7\textwidth]{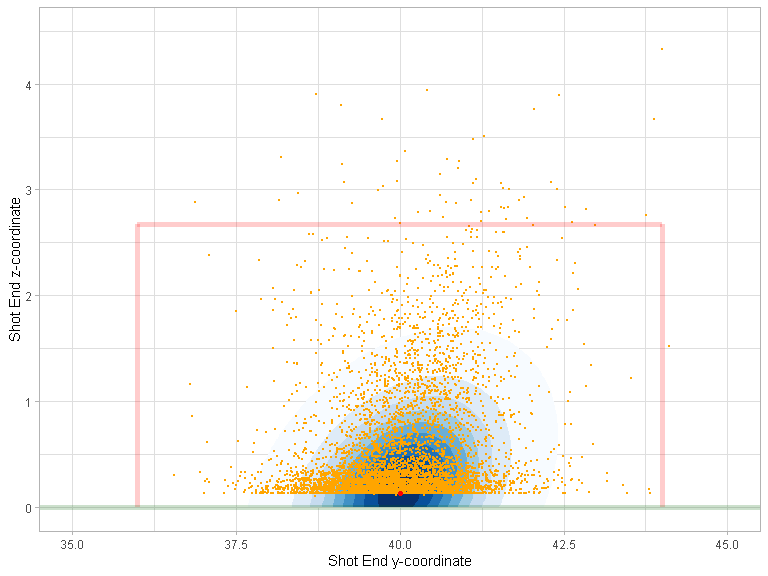}}
\caption[Execution Error Distribution Fits]{Bivariate Gaussian fits for shots aimed at the high (top) and low (bottom) red intentions, based on data from \citet{Hunter}. The cut-off at 0.2 for the end z-coordinates is a result of the computer vision process used by \cite{Hunter} to compute shot end coordinates.}
\label{HunterBivariates}
\end{figure}

We fix the component covariances in our saturated mixture model for computational tractability purposes. To inform our choice of $\mathbf{S}_j$ for each location $\mathbf{m}_j$, we fit two bivariate Gaussian distributions to the empirical data in \citet{Hunter}, one for shots aimed 1.75 yards above the ground and another for shows aimed 0.14 yards above the ground. The end $y$-coordinates of left-footed shots were reflected prior to fitting, since left-footed shots display a symmetric pattern to right-footed shots with respect to the orientation of the principal axis. The fitted covariance matrices for these two intended shot locations are:
\begin{align*}
  \mathbf{S}_{(0.14\textrm{yd})} &= \begin{bmatrix} 0.704 & 0.157 \\ 0.157 & 0.297 \end{bmatrix} &
  \mathbf{S}_{(1.75\textrm{yd})} &= \begin{bmatrix} 0.782 & 0.442 \\ 0.442 & 0.742 \end{bmatrix}
\end{align*}
Figure \ref{HunterBivariates} shows the resulting bivariate Gaussian distributions.

\blue{We assume that the shape of the execution error $\mathbf{S}$ changes linearly with respect to the intended shot height, but is unchanged as the intended location moves laterally.}

\subsection{\blue{Sensitivity Analysis with Respect to Data Preprocessing Choices}}
\label{SensitivityAnalysis}

\blue{In our data pre-processing, we filter the data to only include shots taken from a distance of at least 6 yards. Additionally, we reflect the end $y$ coordinates of left-footed shots in an attempt to exploit symmetry in shooting patterns.}

\begin{figure}[!th]
    \centering
    \subfloat[\centering Original Shot End Coordinates]{\includegraphics[width=0.75\textwidth]{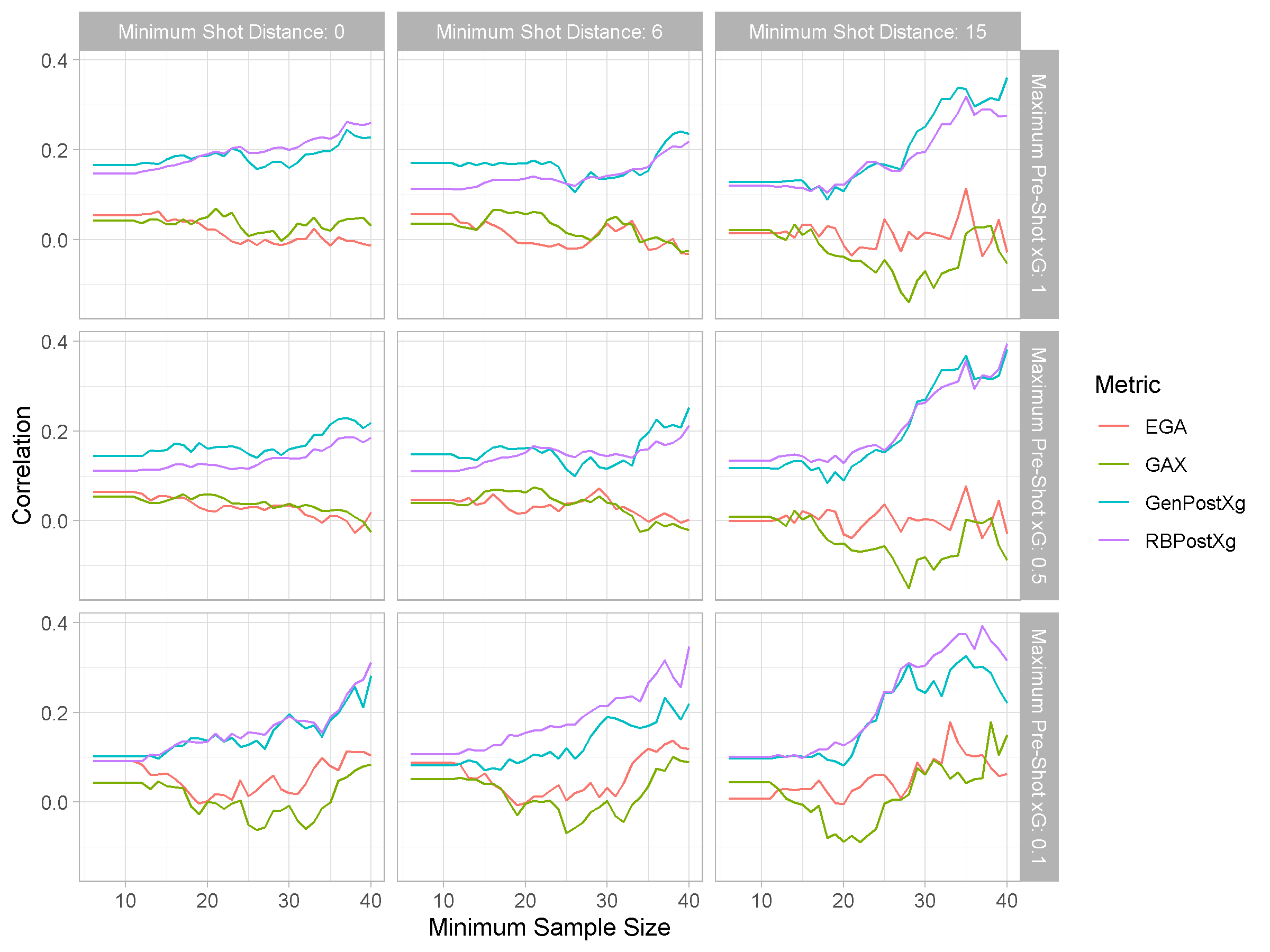}}
    \qquad
    \subfloat[\centering Left-footed Shots Reflected]{\includegraphics[width=0.75\textwidth]{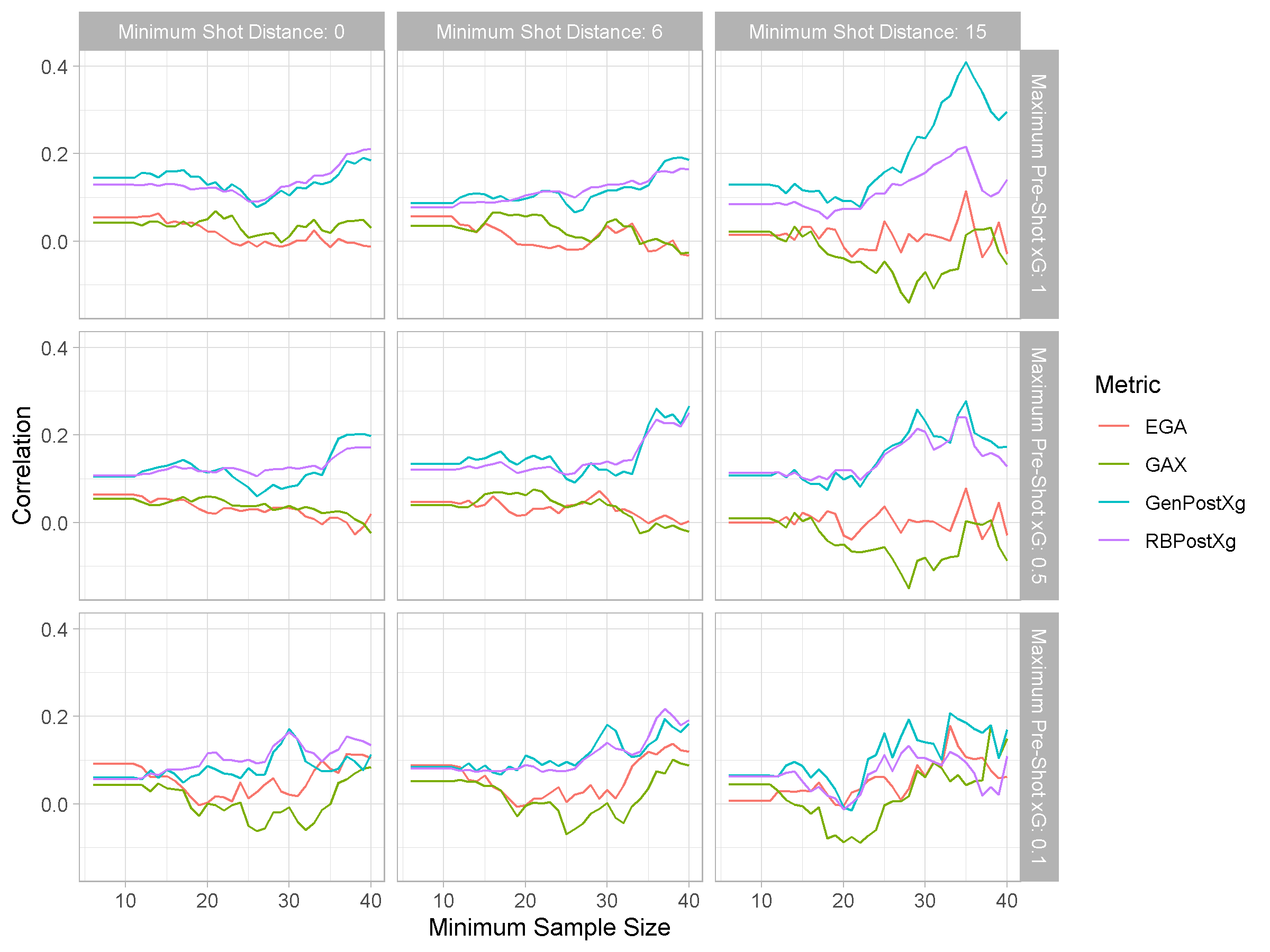}}
    \caption[Sensitivity Analysis]{\blue{Comparison of results under various data processing settings.}}
    \label{SensitivtyAnalysisPlots}
\end{figure}

\blue{To show that our results are robust to the choice of filtering thresholds and choice to reflect left-footed shots, we repeat our analysis under various data processing settings. Specifically, we generate simplified versions of Figure \ref{InterseasonCorrelationsPlot} for nine combinations of filtering thresholds, with or without reflecting left-footed shots. The results are presented in Figure \ref{SensitivtyAnalysisPlots}.}

\blue{Based on these results, we see that our proposed metrics outperform the benchmark metrics for nearly all thresholds and conditions. Additionally, we see that our metrics tend to outperform the benchmark metrics more significantly when a larger set of shots is included in the analysis. This figure confirms that our findings are generally robust to the data processing choices made in our analysis.}